\documentstyle[11pt,aaspp4,flushrt]{article}

\def\hst{{\em HST}}
\def\hnot{H$_0$}
\def\qnot{q$_0$}
\def\msun{\ifmmode {\rm M_\odot} \else M$_\odot$\fi}

\def\msunyr{\ifmmode {\rm M_\odot~yr^{-1}}\else${\rm M_\odot~yr^{-1}}$\fi}
\def\lam{\ifmmode {\lambda} \else {$\lambda$} \fi}
\def\lalpha{L$\alpha$}
\def\muobs{\ifmmode {\mu_{obs}} \else  $\mu_{obs}$ \fi}
\def\mdoto{\ifmmode {\dot{M}_0} \else  $\dot{M}_0$ \fi}
\def\teff{\ifmmode {T_{eff}} \else $T_{eff}$ \fi}
\def\ilam{\ifmmode {I_\lambda} \else  $I_\lambda$ \fi}
\def\inu{\ifmmode {I_\nu} \else  $I_\nu$ \fi}
\def\fnu{\ifmmode {F_\nu} \else  $F_\nu$ \fi}
\def\tauh{\ifmmode {\tau_{\rm H}} \else $\tau_{\rm H}$ \fi}
\def\cm{\ifmmode {\rm cm} \else  cm \fi}
\def\cmmitwo{\ifmmode \rm cm^{-2} \else $\rm cm^{-2}$\fi}
\def\cmmithree{\ifmmode \rm cm^{-3} \else $\rm cm^{-3}$\fi}
\def\cmps{\ifmmode \rm cm~s^{-1}\else $\rm cm~s^{-1}$\fi}
\def\cmpsps{\ifmmode \rm cm~s^{-2}\else $\rm cm~s^{-2}$\fi}
\def\kmps{\ifmmode \rm km~s^{-1}\else $\rm km~s^{-1}$\fi}
\def\kmpspmpc{\ifmmode \rm km~s^{-1}~Mpc^{-1} \else
    $\rm km~s^{-1}~Mpc^{-1}$\fi}
\def\ergps{\ifmmode \rm erg~s^{-1} \else $\rm erg~s^{-1}$ \fi}
\def\ergpspcm{\ifmmode \rm erg~s^{-1}~cm^{-2} 
    \else $\rm erg~s^{-1}~cm^{-2}$ \fi}
\def\ergpspcmphz{\ifmmode \rm erg~s^{-1}~cm^{-2}~Hz^{-1} \else $\rm
   erg~s^{-1}~cm^{-2}~Hz^{-1}$ \fi}
\def\ergpspcmpa{\ifmmode \rm erg~s^{-1}~cm^{-2}~\AA^{-1} \else $\rm
erg~s^{-1}~cm^{-2}~\AA^{-1}$ \fi}
\def\ergpsphz{\ifmmode \rm erg s^{-1} Hz^{-1} \else 
   $\rm erg s^{-1} Hz^{-1}$ \fi}
\def\mdoto{\ifmmode \dot M_0 \else $\dot M_0$ \fi}
\def\eg{e.g.}

\def\cf{cf.}
\def\etal{~et al.}

\received{}
\accepted{}
\journalid{}{}
\articleid{}{}

\slugcomment{}

\begin{document}

\title{The Lyman Continuum Polarization Rise in the QSO PG~1222+228\footnote
{Accepted for publication in Publications of the Astronomical
Society of the Pacific, 2000 May}}

\author{Gregory A. Shields}
\affil{Department of Astronomy, University of Texas, Austin, TX 78712}
\authoremail{shields@astro.as.utexas.edu}

\lefthead{G. A. Shields}
\righthead{Polarization of PG 1222+228}

\begin{abstract}

Some QSOs show an abrupt, strong rise in polarization 
at  rest wavelength $\sim750$~\AA.  If this arises
in the atmosphere of an accretion disk around a supermassive black hole, it may
have diagnostic value.  In PG 1222+228, the polarization rise occurs at the
wavelength of a sharp drop in flux.  
We examine and reject interpretations of this
feature involving a high velocity outflow.  The observations agree with a model
involving several intervening Lyman limit
systems, two of which happen to coincide with the
Lyman continuum polarization rise.  After correction for the Lyman limit
absorption, the continuum shortward of $\lambda 912$ is consistent with a
typical power-law slope, $\alpha \approx -1.8.$  This violates the apparent
pattern for the Lyman limit polarization rises 
to occur only in ``candidate Lyman
edge QSOs''.  The
corrected, polarized flux rises strongly at the wavelength of the polarization
rise, resembling the case of PG 1630+377.  The rise in polarized flux places
especially stringent requirements on models.

\end{abstract}


\section{INTRODUCTION}

Spectropolarimetric observations with the 
{\em Hubble Space Telescope} (\hst) have
revealed an unexpected rise in linear polarization in several QSOs (see review
by Koratkar \& Blaes 1999).  These are radio quiet ``candidate Lyman edge
QSOs'', in which the continuum flux drops rather rapidly at rest wavelengths
$\lambda < 1000$~\AA\  
(Antonucci, Kinney, \& Ford 1989; Koratkar, Kinney \& Bohlin
1992).  Models of accretion disk atmospheres predicted a reduced polarization
in the Lyman continuum because of a diminished contribution 
of electron scattering
to the opacity (Laor, Netzer, \& Piran 1990).  With this
motivation, Impey
\etal\ (1995) and Koratkar \etal\ (1995) used the Faint Object Spectrograph
(FOS) on \hst\  to obtain ultraviolet spectropolarimetry of several QSOs with
redshifts sufficient to bring the Lyman continuum within the observed
wavelength band.  The surprising result, in several cases, was a rapid {\em
rise} in polarization in the Lyman continuum.  From values $\lesssim 1$ percent
in the optical and near ultraviolet, the observed polarization rises around
rest wavelength 750~\AA\ to values $\sim5$ percent 
in several objects, and to $\sim20$
percent in PG 1630+377.  

This phenomenon has inspired several attemps at
explanation.  Blaes and Agol (1996) found that, for 
effective temperatures $\teff \approx 25,000$~K and low effective gravities, a
polarization rise of up to
$\sim5$~percent at about the observed wavelength could occur naturally in QSO
disk atmospheres.  This results from the interplay of electron scattering,
bound-free opacity, and the temperature gradient in the atmosphere.  However,
Shields, Wobus, and Husfeld (1998, hereinafter SWH) showed that  the effects of
the relativistic transfer function destroy the agreement between this model and
observation.  Beloborodov and Poutanen (1999) suggested a model involving
Compton scattering in a corona or wind, but this model appears to have trouble
giving the rapid rise in polarized flux observed in PG 1630+377 (Blaes \&
Shields 1999).  Lee and Blandford (1997) 
discussed the possible role of scattering by
resonance lines of heavy elements (see Section 5).

SWH showed that, if
the polarization is assumed to rise sharply at 
$\lambda 912$ in the rest frame of
the orbiting gas, then relativistic effects would 
naturally produce the wavelength
dependence of the observed polarization.  This may offer a way of
measuring the black hole spin, but the physical mechanism for the polarization
rise remains unknown.

PG 1222+228 is a $\rm B \approx 15.5$ radio quiet 
QSO (Schmidt \& Green 1983) whose
polarization rise at
$\lambda \approx 750$~\AA\ coincides with a sharp drop 
in flux (Fig. 1, 2).  Impey
\etal\ (1995) noted this and attributed 
it to a coincidental Lyman limit system (LLS),
corresponding to an identified absorption line system at 
z = 1.486.  However, the
coincidence of a broad absorption feature with a polarization rise also is
observed for broad absorption line (BAL) QSOs.  In these objects, outflowing
gas at velocities $\sim10^4~\kmps$ produces blueshifted absorption troughs,
typically seen in the resonance lines of H I, C IV, N V, O VI, Si IV, and
sometimes Mg II (Weymann \etal\ 1991; 
Arav, Shlosman, \& Weymann 1997). Spectropolarimetric
observations (\eg, Ogle 1997; Schmidt \& Hines 1999; 
Ogle \etal\ 1999) often show a rise in
polarization in the troughs, reaching values as 
high as $\sim8$ to 10 percent from $\sim1$
percent at unabsorbed wavelengths.  This is explained 
in terms of scattering of some of
the continuum by an extended region that is 
not covered by the BAL flow (Hines \& Wills
1995; Goodrich \& Miller 1995; Cohen \etal\ 1995). This pattern resembles the
polarization rise and flux drop in PG 1222+228.

This paper addresses two questions: (1)  Does the polarization rise in PG
1222+228 result from an intrinsic absorber, analagous to the situation in the
BAL QSOs?  (2)  If the 
drop in flux in PG 1222+228 is an intervening LLS, what are
the consequences of correcting the observed, polarized continuum for this
absorption?

\section{INTRINSIC ABSORPTION IN PG 1222+228?}

We first consider the possibility that the flux drop and coincidental
polarization rise in PG 1222+228 results from some kind of intrinsic
absorption.  Two possibilities, considered below, are that 
it is a BAL outflow, or that it
is an unusual, intrinsic LLS.  

In either case, one issue is
the behavior of the polarized flux as a function of wavelength.  As discussed above, if the
polarization rise results from the selective absorption of the directly viewed continuum but
not the scattered continuum, one might expect a smaller drop (but generally not a rise) in
the polarized flux, 
$I_p = p\ilam$.  In order to examine this, we have
rebinned the data of Impey \etal\ (1995), kindly made available in reduced form
by C. Impey and C. Petri (1999).  These data consist of a spectrum with the G190H
grating covering $\lambda\lambda1575$ to 2320 at 0.37~\AA\ per pixel, and
a spectrum with the G270H grating covering the range 2224 to 3295 at 0.52~\AA\
per pixel.  The G190H data shortward of $\lambda1994$ have low signal-to-noise and were
not presented by Impey \etal\ (1995). We used seven wavelength bins (in \AA):
(1)1994--2224, (2) 2224--2287,  (3) 2287--2319, (4) 2319--2492, (5)
2492--2761, (6) 2761--3029, (7) 3029--3295.  Bins 2 and 3
involve an average of the two overlapping spectra, and bin 3 is a narrow bin containing
the flux drop at $\lambda 2300$.  The resulting values of
\ilam, $q \equiv Q/I$ and $u \equiv U/I$ are tabulated in Table 1, along with the
polarization, $ p = \sqrt{q^2 + u^2}$, and its position angle, $\theta$.  For the
polarized flux, we use the rotated Stokes flux
$Q^\prime$ and polarization
$q^\prime \equiv Q^\prime/I$ (\cf\  Koratkar etal\ 1995), referred to a position angle of
168 degrees. 
 This is based on the mean polarization position angle of the HST data,
which is in reasonable agreement with optical observations (Stockman \etal\ 1984; Webb
\etal\ 1993).   (The use of $q^\prime$ is appropriate if the position angle of the
polarization is constant with wavelength.  Table 1 supports this and also shows that the
polarization $p$ is reasonably consistent with $q^\prime$.)
 These quantities are plotted in Figure 2.
The shortest wavelength
bin has larger polarized flux than the longer wavelength bins.  At face value, this would
weigh against an intrinsic absorber model for PG 1222+228; but it involves a single
wavelength bin with substantial error bars. Therefore, we consider other aspects of the
two outflow models.

\subsection{BAL Absorption}

The rest wavelength of the onset of the absorption feature in PG 1222+228 is
$~\sim750$~\AA.  Some BAL QSOs show absorption by Ne VIII $\lambda$775 (\eg,
Arav \etal\ 1999; Telfer \etal\ 1999).  This might be a candidate
for the feature in question, in as much as BALs often set in at a wavelength
somewhat blueshifted from the emission-line redshift.  However, Ne VIII normally
is accompanied by absorption in O VI
$\lambda 1035$, N V $\lambda 1240$, and C IV $\lambda 1550$.  
There is no indication of broad
C IV or Mg II absorption in the spectrum of PG 1222+228 (Sargent, Steidel, \& Boksenberg
1988; Steidel \& Sargent 1992).  The
\hst\ spectrum shows a shallow trough at $\lambda 3000$ to $\lambda 3070$ that could be a
weak O VI feature; but this may simply be a cluster of lines, including several
strong
\lalpha\ lines indentified by Impey \etal\ (1996).  Photoionization models by
Hamann (1997) indicate a range of ionization parameters for which the
fractional abundance of Ne$^{+7}$ exceeds that of O$^{+5}.$  However, given the normal
ratio of oxygen to neon abundances, the O VI feature would likely be strong in
a situation giving strong Ne VIII.  

The flux drop at
$\lambda 750$ does not recover, with decreasing wavelength, in a way
suggestive of a BAL (Figure 1).  The spectrum has not fully recovered by rest wavelength
650~\AA, corresponding to an an
outflow velocity of more than 40,000 \kmps\ if attributed to Ne VIII.  Some moderately
narrow ``mini-BAL'' features have been observed at such high velocities (Hamann \etal\ 
1997), but true BAL troughs rarely reach such velocities.  The same can be said in
connection with the possibility that the $\lambda 750$ feature corresponds to a blend of
features including N III, N IV, O IV, S VI, and Ne VIII seen in BAL QSO spectra at this
wavelength (\eg, Arav \etal\ 1999).  Moreover, given the range of ionization
stages contributing to this blend, C IV and Si IV absorption 
would likely accompany it.

Recent work has shown that BAL QSOs systematically have weak soft X-ray
emission. BAL QSOs have optical to X-ray slopes $\alpha_{ox} \lesssim -2.0$, whereas
nonBAL QSOs tend to have $\alpha_{ox}$ in the range -1.3 to -1.8 (Brandt, Laor,
and Wills 1999). Here, $\alpha_{ox}$ is defined by $F_x/F_o =
(\nu_x/\nu_o)^{\alpha_{ox}}$, where $F_x$ and $F_o$ are the flux densities ($\fnu$)
at 2 keV and 3000~\AA, respectively.  ROSAT pointing observations give a flux of
$6\times10^{-14}~\ergpspcm$  for PG 1222+228 at a significance level of
$3.3\sigma$ (Mushotzky 1999).  If we assume a ``normal'' power-law slope of
$\alpha = -1.6$ over the 0.2 to 2 keV ROSAT band (Brandt \etal\ 1999), we find 
$F_x = 1.5\times10^{-31}~\ergpspcmphz$ at 2 keV rest energy.  Optical
spectrophotometry (Wampler \& Ponz 1985; Bechtold \etal\ 1984) implies $F_o \approx
1.1\times10^{-26}\ \ergpspcmphz$ at rest wavelength 3000~\AA.  (These are
observed fluxes at the wavelength corresponding to the indicated rest
wavelength.)  From this, we find $\alpha_{ox} = -1.8$ for PG 1222+228.  This
result is uncertain because of the marginal X-ray detection and the
possibility of variability, but at face value it is more consistent with a
non-BAL than a BAL QSO.  (Wilkes \etal\ 1994 quote an uncertain value 
$\alpha_{ox}
\approx -1.8$ from Einstein data.)

We conclude that the flux drop at $\lambda 750$ in PG 1222+228 is unlikely to be a BAL
feature.

\subsection{An Instrinsic Lyman Limit System?}

Impey \etal\ (1995) suggested that the flux drop at $\lambda 750$ was an intervening LLS. 
We show below that Lyman limit absorption does indeed give a good fit to the spectrum. 
This fit, however, leaves open the question of the location of the absorbing gas. 
Because of the coincidence with the polarization rise, we consider here the
possibility that the feature is an {\em intrinsic} LLS, associated with a high
velocity outflow from the QSO.  Impey
\etal\ (1996) identify \lalpha\ absorption lines at z = 1.4857, 1.5238, 1.5272,
and 1.5650 that might be associated with the $\lambda 750$ feature, if it is taken to
be a LLS.  A redshift of 1.486 corresponds to an outflow velocity $\sim60,000\
\kmps.$  As noted above, this is not unprecedented for a QSO outflow producing
absorption lines.  However, the lines associated with the redshift systems in
question in PG 1222+228 are narrow, 
and narrow lines with relative velocities greater than 5000
\kmps\  usually are assumed to be intervening.  For a Lyman edge optical depth of
unity, the measured equivalent widths of the
\lalpha\ lines are consistent with a Doppler parameter
$b \lesssim 30~\kmps$ (see below),  normal for
an intervening LLS. In contrast, the mini-BALs observed at such
high velocities have widths of order $1000~\kmps$ (Hamann \etal\ 1997).  
Could the feature in PG 1222+228 nevertheless be caused by ejected
material with a high outflow velocity and a small velocity dispersion?

 We are not aware of any other case in which a high velocity LLS with narrow lines
has been shown to be intrinsic. 
However, the existence of many intrinsic, narrow, high velocity absorption systems in
QSOs has been proposed by Richards \etal\ (1999).  These authors
compare the incidence of absorption-line systems, per unit relative outflow
velocity, for highly luminous QSOs with that for less luminous ones.  In
the velocity range 5000 to 75,000 \kmps, they find that absorption systems
have a substantially larger frequency in luminous systems.  Since intervening
systems should have no dependence on QSO luminosity (assuming that discovery
systematics are accounted for), Richards \etal\ conclude that at least
the excess number of systems in the high luminosity QSOs are intrinsic.

A peculiar absorption line ratio in the z = 1.94 system in PG 1222+228 has
been noted by Ganguly \etal\ (1998).  These authors present high resolution
spectra that show two narrow components, separated by $\sim17~\kmps$,
with very different strengths of the Al II and Al III absorption
lines.  Photoionization models
indicate that the component with strong aluminum lines must have an anomalously
high abundance of aluminum.  This is reminiscent of claims of
unusual abundances in BAL QSOs, including an excess of aluminum (\eg, Turnshek
\etal\ 1996; Junkkarinen \etal\ 1997; Shields 1997).  The reality of these
abundance anomalies is in doubt, because of the effects of partial covering of
the continuum source  (Arav 1997).  However, the basic observation of
anomalously strong Al lines may be a possible parallel between PG 1222+228 and
the BAL QSOs, where outflowing gas is clearly present.  If this is a hint that
the narrow, z = 1.94 system may be intrinsic, perhaps it adds plausibility to
the idea that the z = 1.486 system (or its neighbors) may also be intrinsic.

What might be the geometry of an intrinsic LLS in PG 1222+228?  In order to
explain the polarization rise, the absorber would have to intercept the line of
sight to the continuum source but not the scattering source.  The latter is
often attributed to a wind driven off the inner edge of a ``dust torus'' (Krolik \&
Begelman 1986).  The location of this may be related to the dust sublimation radius,
$\sim0.2 L_{46}^{1/2}$ pc, where $L_{46}$ is the bolometric luminosity of the central
source in units $10^{46}~\ergps$ (Laor \& Draine 1993). The absorbing material, at a velocity
of
$\sim0.2c,$ would be at a smaller radius in order not to cover the scattering source. Its
high velocity suggests an origin at a small radius where the escape velocity is
of order the observed outflow velocity.  The escape velocity from a central mass of
$M$ is 0.2$c$ at a radius
$10^{15.6} M_9$~cm, where $ M_9 \equiv M/10^9~\msun$.  Let us assume the absorbing
material is at a radius of $\lesssim10^{18}$~cm, large enough to obscure the ultraviolet
emitting part of an accretion disk but not the scattering source.  At the observed
speed, the crossing time would be a few years or less.  Thus, the material should
change radius substantially in the seven years between the Sargent \etal\ (1988) and
the Impey \etal\ (1996) observations, and one might expect some change of velocity.  These
authors, however, quote velocities for the
$z = 1.486$ and 1.524 systems that agree within $\Delta z \approx
0.001$.  This corresponds to a change in outflow velocity of less than
$\sim100~\kmps$. Such precise stability seems difficult to achieve. (Note,
however, the stability of narrow features within the BAL
profiles of some QSOs [\cf, Weymann 1997]).  A further problem  involves the
narrowness of the absorption lines.  The radius of the ultraviolet emitting part
of the disk would be at least
$\sim10$ gravitational radii, or about $10^{15.1} M_9$~cm.  If the absorber is at a radius
$\lesssim 10^{18}$ cm, then the line of sight through the absorber to
different parts of the continuum source would likely give noticeably different
projected flow velocities. The observed linewidths (see below) are
$\lesssim30~\kmps$, less than one thousandth of the outflow velocity.  

These difficulties with the intrinsic absorber model for PG 1222+228 encourage
us to examine the straightforward idea of an intervening LLS.

\section{INTERVENING ABSORPTION}

Impey \etal\ (1995, 1996) attributed the flux drop at observed wavelength $\lambda 2300$
to a LLS associated with the $z = 1.4857$ absorption line system.  They identify
absorption lines of H I, C III, N I, N II, Si II, and Si III.  They also
identify systems at 1.5238, 1.5272, and 1.5650, which have multiple Lyman
lines and, for $z = 1.5238$, C III.  Sargent, Steidel, \& Boksenberg (1988)
measure C IV $\lambda\lambda 1548$, 1551 in the 1.486 and 1.524 systems with
equivalent widths $\sim0.7$~\AA.  However, Steidel
\& Sargent (1992) give a spectrum showing no detectable Mg II absorption.  

Simple
estimates suggested that the $\lambda 2300$ flux drop might be too gradual to be
attributed to the converging Lyman lines of the $z = 1.486$ system.  Therefore, we
computed a model spectrum involving an assumed power-law continuum and absorption by the
hydrogen lines and bound-free continuum.  The column density of H I was
parameterized by the Lyman edge optical depth, $\tau_H$, and the lines were assumed to
have a Gaussian profile with a Doppler parameter set to a typical value
$b = 30~\kmps.$  (A larger line width would give
an excessive equivalent width for \lalpha\ for the required \tauh.  The observed
decrement of the Lyman line equivalent widths actually suggests a narrower core
line width, with some of the
\lalpha\ equivalent width being attributable to a broader component of modest
column density.  These details do not affect our conclusions.)  Inclusion of 50
Lyman lines proved more than adequate to trace the convergence to the continuum
optical depth.  The model spectrum was convolved with a single component Gaussian
instrumental line profile following the discussion of Impey \etal\ (1996), using a
FWHM of 2.9~\AA\ for G190H and 4.2~\AA\ for G270H.  If only the $z = 1.486$ system
contributes to the LLS, then the Lyman line convergence is too close to the Lyman
limit ($\lambda 912$) to fit the observed feature.  However, allocation of some H I column
density to both the 1.486 and 1.524 systems gave a good fit.  Two additional LLS
appear to be associated with the
$z = 1.938$ and 1.174 systems.  Figure 3 shows the resulting model spectrum, along
with the observed flux, binned in intervals of $\sim7.3$~\AA.  The model has
$\tauh =  (1.0, 0.5, 0.8, 0.6)$ for z = (1.174, 1.486, 1.524, 1.938), respectively.
(The value of \tauh for z = 1.174 is uncertain because of the poorly determined
amount of scattered light below 2000 \AA.) 
The three distinguishable LLS show good agreement with the expected
$\nu^{-3}$ behavior of the Lyman continuum optical depth above threshold, for an intrinsic
continuum slope in the range $\alpha
\approx -1.5$ to -2.0.  This is consistent with the slope -1.8 found by Zheng
\etal\ (1997) for their composite QSO spectrum in the wavlength range $\lambda
600$ to $\lambda 1050$.  A value $\alpha = -1.8$ is assumed in the fit shown in Figure 3. 
The gradual descent of the observed flux toward the Lyman limit for the $z = 1.938$ system is
a puzzle, but it may involve the effects of unrelated absorption lines.  An
understanding of this is important, as it would play a role in the
classification of PG 1222+228 as a candidate Lyman edge QSO.
Observations at higher spectral resolution would help to clarify the situation.

Sargent, Steidel, and Boksenberg (1989) discuss the statistics of LLS in
QSOs.  For the redshift range in question, they give a mean incidence of LLS of $N(z)
\approx 1.5$ per unit redshift.  Their data show many
 QSOs with multiple LLS,
although the number of LLS in PG 1222+228 may be somewhat higher than typical.  However, the
efforts to measure Lyman continuum polarization in QSOs to some extent targeted the
candidate Lyman edge QSOs, and objects with LLS of moderate optical depth may have an
enhanced probability to be included.

We conclude that an intrinsic power-law continuum, together with cosmologically
intervening Lyman limit absorption, provides a straightforward explanation of
the ultraviolet spectrum of PG 1222+228.  

\section{THE INTRINSIC POLARIZED CONTINUUM}

The various LLS in PG 1222+228  substantially attentuate the observed continuum. 
What is the behavior
of the {\em polarized flux} when corrected for the absorption?  In view of the
uncertainties in the measured polarization, a sufficient procedure is
to estimate the polarized flux by multiplying the measured polarization in the chosen
wavelength bins by the assumed intrinsic continuum flux.  For this, we use the same
wavelength bins described earlier, and the $\inu^{PL}
\propto \nu^{-1.8}$ power-law continuum used in our fit.  The resulting rotated
Stokes flux, $Q_{*}^\prime = q^\prime \times \ilam^{PL}$, is shown in Figure 4.  We see that
the Stokes flux now rises strongly with decreasing wavelength in the region of the
polarization rise.  This resembles the result found for PG 1630+377 by Koratkar \etal\
(1995).  A  rising polarized flux is an important constraint on models for the origin of
the polarization rise.

SWH showed that the wavelength dependence of the flux and
polarization in PG 1222+228 and PG 1630+377 could be fit with an ad hoc model
involving an accretion disk.  The disk radiates as a black body, but the brightness
temperature is depressed below the effective temperature for wavelengths below the Lyman
limit,  simulating a Lyman edge in the disk atmosphere.  The polarization is assumed to
rise abruptly at $\lambda 912$ by an arbitrary amount.  Relativistic effects give a
blueshifted, but still fairly abrupt polarization rise in the observed spectrum.  The
models are characterized by $a_*,$ the dimensionless angular momentum of the hole;
the black hole mass; the accretion rate, $\mdoto \equiv {\dot M}/(1~\msunyr)$; and the
viewing angle, $\mu_{obs} = cos(\theta_{obs})$.  SWH found that
$a_* = 0.5$ gave approximately the observed wavelength for the polarization rise, for a
relatively edge-on viewing angle.  Their fits to both objects had a fairly low value of
$T_{max}$, the maximum disk effective temperature, as required by the dropping flux
in the Lyman continuum region.  The correction for LLS absorption in PG 1222+228 hardens
the far ultraviolet spectrum, and a higher value of
$T_{max}$ is required to fit the energy distribution.  Figure 5  shows the
continuum flux and polarization for a model with $a_* = 0.5,$ $M_9 = 8.8$ and $\mdoto =
86$.  (We have assumed \hnot = 70 \kmpspmpc and \qnot = 0.5.)  The model agrees
reasonably well with the corrected flux in the Lyman continuum and with the longer
wavelength measurements.  Although the corrected flux was assumed to be a
power law, a disk continuum would also likely fit the observed flux, given some
freedom to adjust the LLS optical depths.  This model has a step-function rise
in polarization from negligle polarization at wavelengths longward of
$\lambda 912$ to an ad hoc value of 2.1 times the Chandrasekhar (1960) value for
a pure scattering atmosphere. The model predicts an observed
polarization and polarized flux that rise at a wavelength substantially
blueshifted from $\lambda 912$, but the rise is more gradual than observed. This
underscores the need for improved polarization measurements of this object.
If the relativistic transfer function gives too gradual a polarization rise, even
for an instantaneous polarization rise in the rest frame of the orbiting gas, then
accretion disk models for the polarization rise will face a serious problem.

The adopted model parameters give a bolometric luminosity $L/L_E = 0.36,$ where $L_E$ is
the Eddington limit.  Such a high value of $L/L_E$ is barely consistent with a
geometrically thin disk.  A larger value of $a_*$ would allow a larger mass for the
required
$T_{max}$, but then the polarization rise would occur at a wavelength shorter than
observed (\cf\ SWH).  Evidently, an accretion disk fit to the corrected continuum of PG
1222+228, in the manner of SWH, pushes the disk parameters to the limits.  Conceivably,
the thickening of the disk corresponding to the large value of $L/L_E$ may be related to
the origin of the polarization rise.

\section{DISCUSSION}

The Lyman continuum polarization rises are among the more puzzling recent
observational discoveries concerning QSOs.  The wavelength dependence, rising rather
abruptly at nearly the same rest wavelength in the several known cases, suggests a
connection with the bound levels of atoms.  The proximity of the feature to
$\lambda 912$ further suggests an association with the Lyman edge of hydrogen.  The ad hoc
model of SWH supports an association with the Lyman edge and raises the possiblity of
confirming the presence of a relativistic disk and constraining its parameters.
However, attempts to fit the feature with a physical model have encountered 
difficulties.  This is an important problem for QSO theory.

Are the reported polarization rises real?  The
coincidence of the polarization rise in PG 1222+228 with a sharp drop in flux might
raise the question of background problems with the FOS spectropolarimeter.  Impey \etal\
(1995) argue that the polarization is unlikely to be less than 2.7\% around 2000 \AA\ under
any reasonable assumption for the FOS background.  However, the degree of polarization in
the far ultraviolet is uncertain by at least a factor two because of systematic errors
involving background and scattered light in the FOS. The observed polarization rises in
several QSOs occur at different observed wavelengths but similar rest wavelengths.  We are
not aware of any polarization rises of this nature in FOS spectropolarimetry of BL Lac
objects or stars.  There is an urgent need for a renewed capability for ultraviolet
spectropolarimetry from space to confirm and extend the measurements.

Lyman continuum polarization rises have heretofore been associated with the candidate
Lyman edge QSOs (see discussion by Koratkar \etal\ 1998).  Our results suggest that PG
1222+228 may not be a true member of this class.  
Measurements of the Lyman continuum polarization in additional QSOs are needed to
clarify the frequency of occurence of the phenomenon and to look for correlations with
features in the continuum flux, in the line intensities and polarization, and other
properties.  Observations to shorter rest wavelengths are needed to determine whether
the polarization falls or continues to rise.  Measurements of the time dependence of the
polarization rises would be most interesting.  The emitting
radius of an accretion disk would be light weeks.

Lee and Blandford (1995) considered a model for the far ultraviolet polarization 
rise of QSOs that did not involve the Lyman edge.  Noting that a number of resonance
lines of heavy elements fall in the rest wavelength range where the polarization rises, they
suggested that resonance scattering of the QSO continuum might produce the observed
polarization.  Such a model could produce a rising polarized flux, since the polarization
could be essentially zero at wavelengths without scattering contributions.  As noted above,
the polarization rise in PG 1222+228 may be too steep for models involving an accretion
disk.  In this case, alternative models such as resonance scattering may hold promise.  We
note that the polarized flux spectrum of PG 1630+377 (Koratkar
\etal\ 1995) shows a strong rise at the wavelength of the N V emission line.

The claim by Richards \etal\ (1999) that luminous QSOs have many
intrinsic, narrow, high velocity C IV absorption systems has important implications. 
This would complicate the use of such systems to probe the evolution of galaxies and the
intergalactic medium.  The mechanism for producing the absorbing clouds would add
another challenge to the subject of outflows from QSOs.  Surveys of QSOs at
different luminosities, to a uniform standard of signal-to-noise ratio, would allow
confirmation of the claimed higher incidence of absorption in more luminous QSOs. 
Tests of the intrinsic nature of narrow absorptions have been summarized by Barlow,
Hamann, \& Sargent (1997).  These include time variability of the depth and profile of the
lines and evidence for saturated but nonblack line profiles.  Detection of changes
in the velocities of the lines would be most revealing.  Chemical abundances
may be an indicator, given evidence for high abundances in the broad absorption and
emission-line regions of QSOs (\eg, Hamann \& Ferland 1993).  If C IV systems in high and
low luminosity QSOs have similar, mostly subsolar abundances, this might argue for their
intervening nature.

\acknowledgments

The author is grateful to C. Impey and C. Petri for providing advice and the
reduced HST observations and to C. Sneden for the use of a computer subroutine. 
The work has benefitted from discussions and communications with  R. Antonucci, O. Blaes,
R. Blandford, R. Ganguly, M. Malkan, R. Mushotzky, G. Richards, R. Weymann, and B. Wills.
This material is based in part upon work supported by the Space Telescope Science
Institute under Grant No. GO-07359.02.  This work was carried out in part at the Institute
for Theoretical Physics, University of California, Santa Barbara, supported by NSF grant
PH94-07194.

\clearpage

\centerline{{\bf Captions for Figures}}


\figcaption[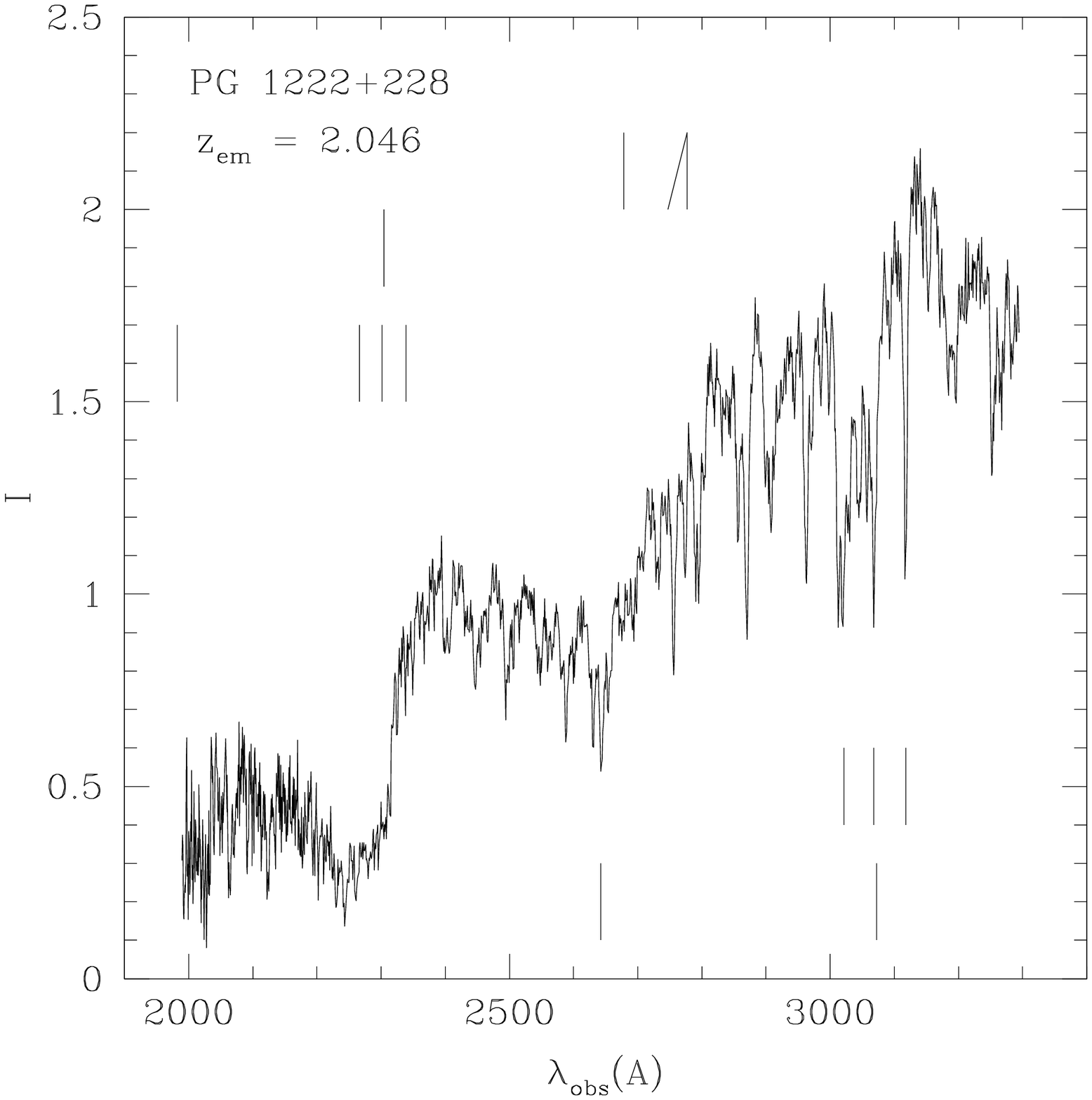]{Ultraviolet spectrum of PG 1222+228 observed with \hst\ by
Impey \etal\ (1995, 1996). Figure gives flux at earth in $10^{-15}~\ergpspcmpa$
as a function of observed
wavelength.  Vertical lines give the positions of the Lyman limit (above)
and \lalpha\ (below) for redshifts z = 1.174,
1.486, 1.524, 1.527, 1.565, and 1.938.  Also shown is the 
Lyman limit for the emission-line redshift of 2.046. Data plotted here are from the G190H
spectrum for $\lambda < 2224$ and G270H for longer wavelengths (see text).  
Data is binned by 2 pixels.
Data courtesy of C. Impey and C. Petri (1999). 
\label{fig1}}

\figcaption[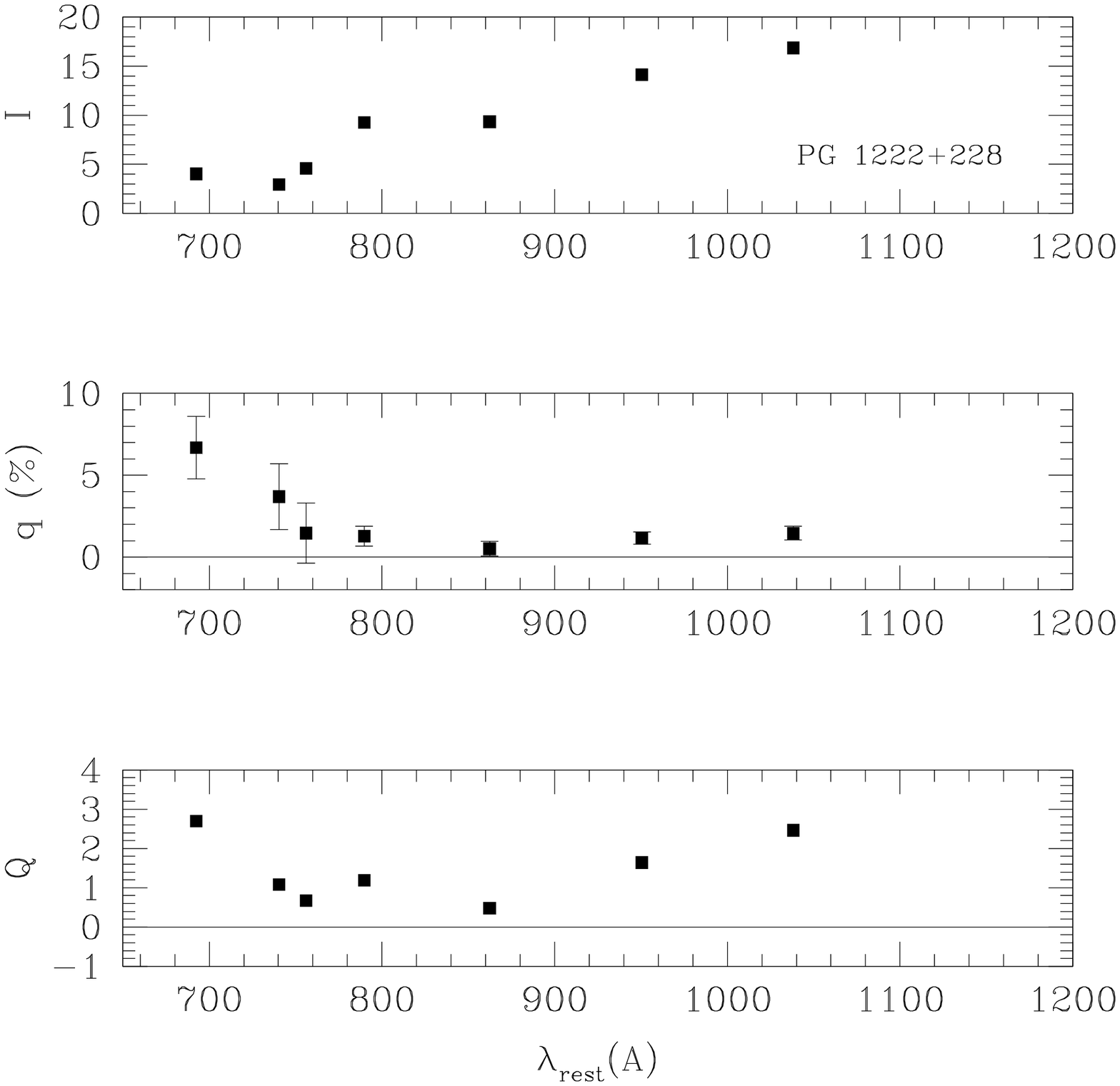]{Flux and polarization of PG 1222+228 from \hst\ observations
by Impey \etal\ (1995, 1996), binned as described in the text.  Abscissa is
rest wavelength in terms of the emission-line redshift.  
$I$ is the measured flux in units of $10^{-16}~\ergpspcmpa$.  
Rotated Stokes flux $Q$, called
$Q^\prime$ in the text, is referred to the mean position angle of 168
degrees.  The fractional rotated Stokes parameter, $q = Q/I,$ is given as a percentage.
\label{fig2}}

\figcaption[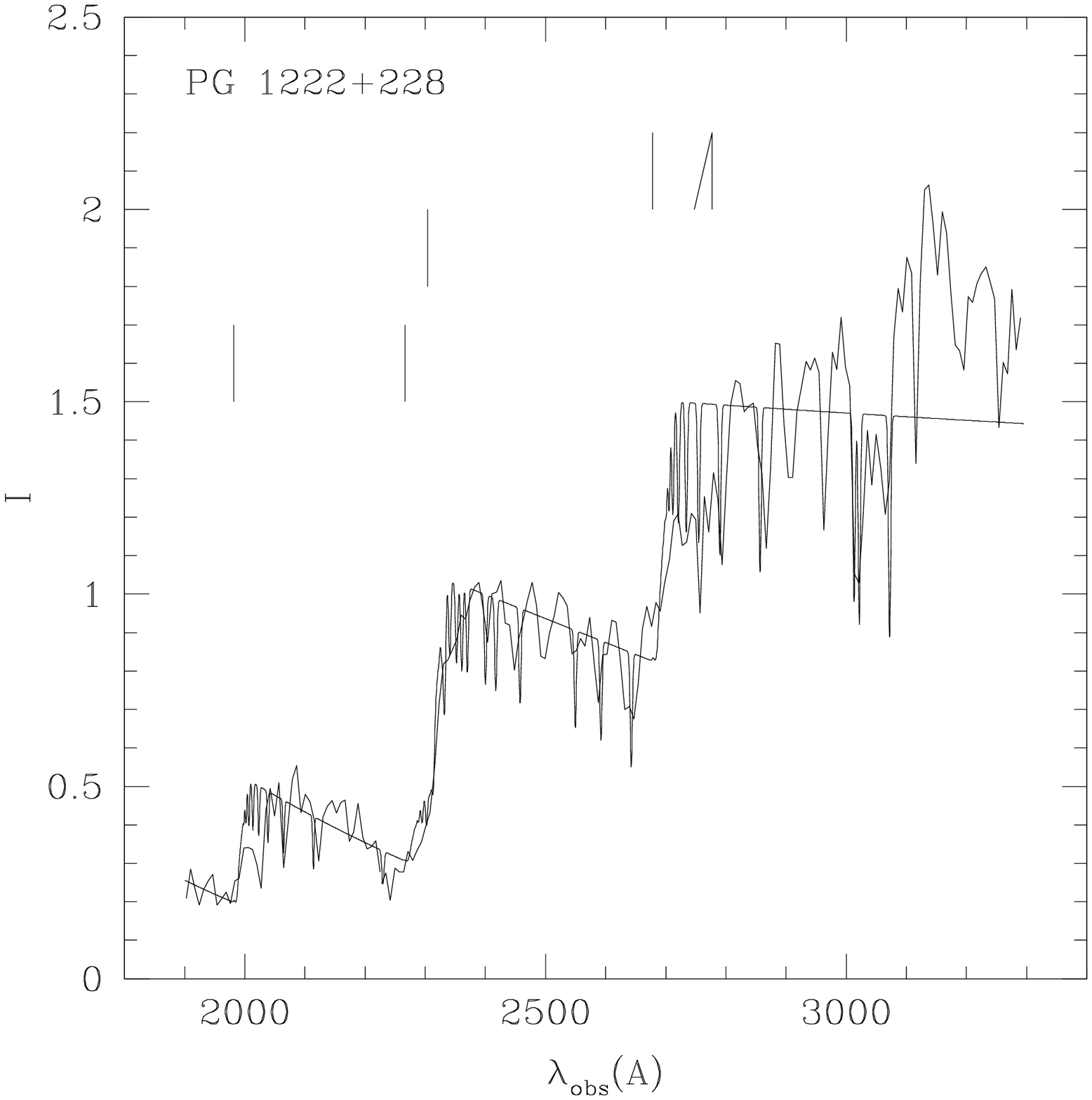]{Observed flux of PG 1222+228 compared with model involving
power-law continuum and absorption by H I Lyman lines and continuum at four
redshifts as described in the text.  Observations have been binned into intervals
of approximately 7.3~\AA\ for clarity.  Axes are same as in Fig. 1
\label{fig3}}

\figcaption[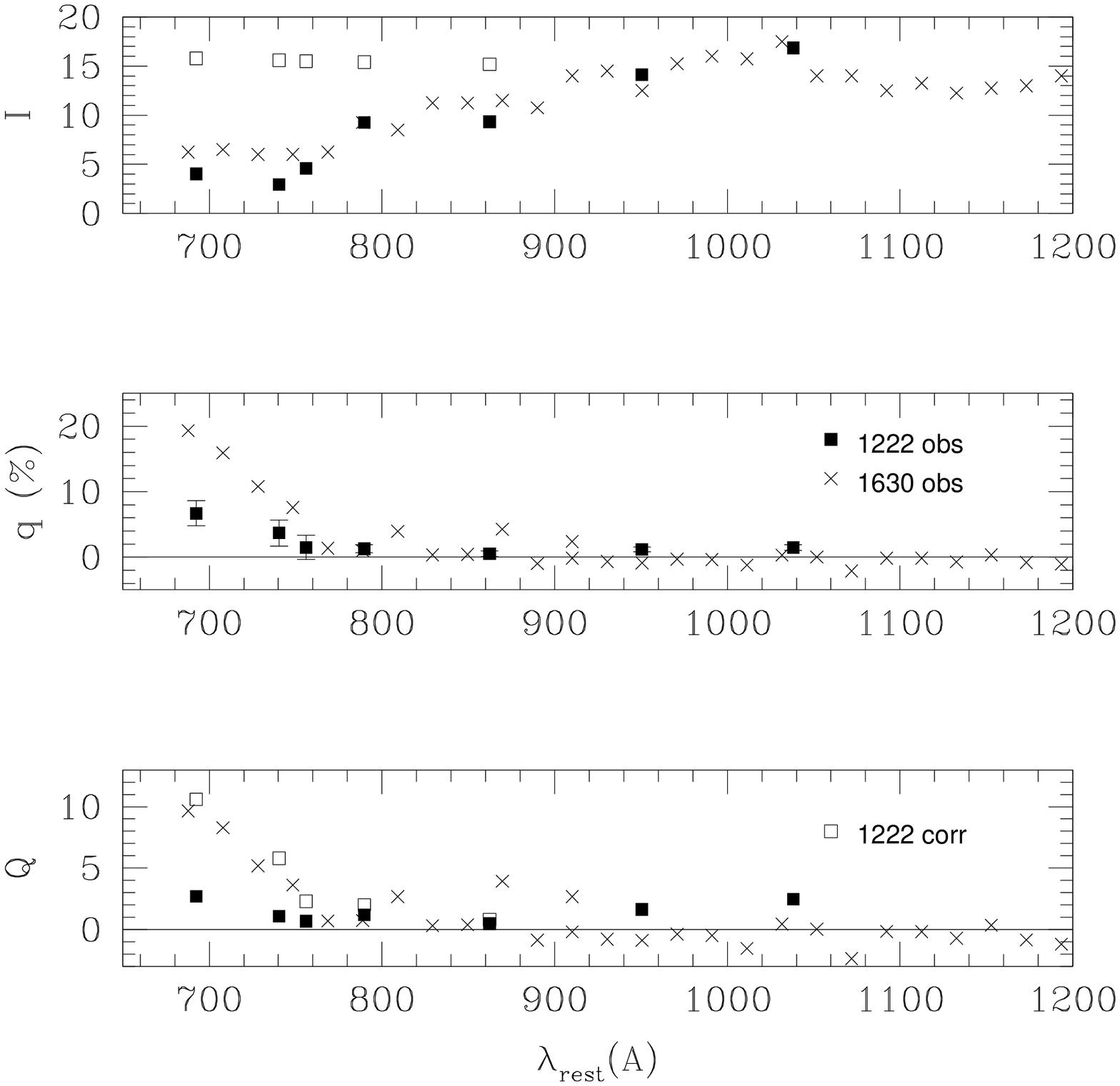]{Flux and rotated Stokes flux of PG 1222+228 before 
(filled squares) and after (open squares)
correction for absorption by Lyman limit systems as described in the text.  Also
shown (crosses) are observations of PG 1630+377 by Koratkar \etal\ (1995).  
For PG 1630+377, the Stokes flux has been rotated to the mean position angle of 127
degrees, following Koratkar \etal\  The plotted values of I and Q have been scaled to
correspond to the peak values in PG 1222+228 in order to emphasize the wavelength
dependences.  The corrected, polarized flux of PG 1222+228  
($Q_*^\prime$ in the text) rises strongly below
$\lambda 750$, resembling the case of PG 1630+377.  Axes are same as in Fig. 2.
\label{fig4}}

\figcaption[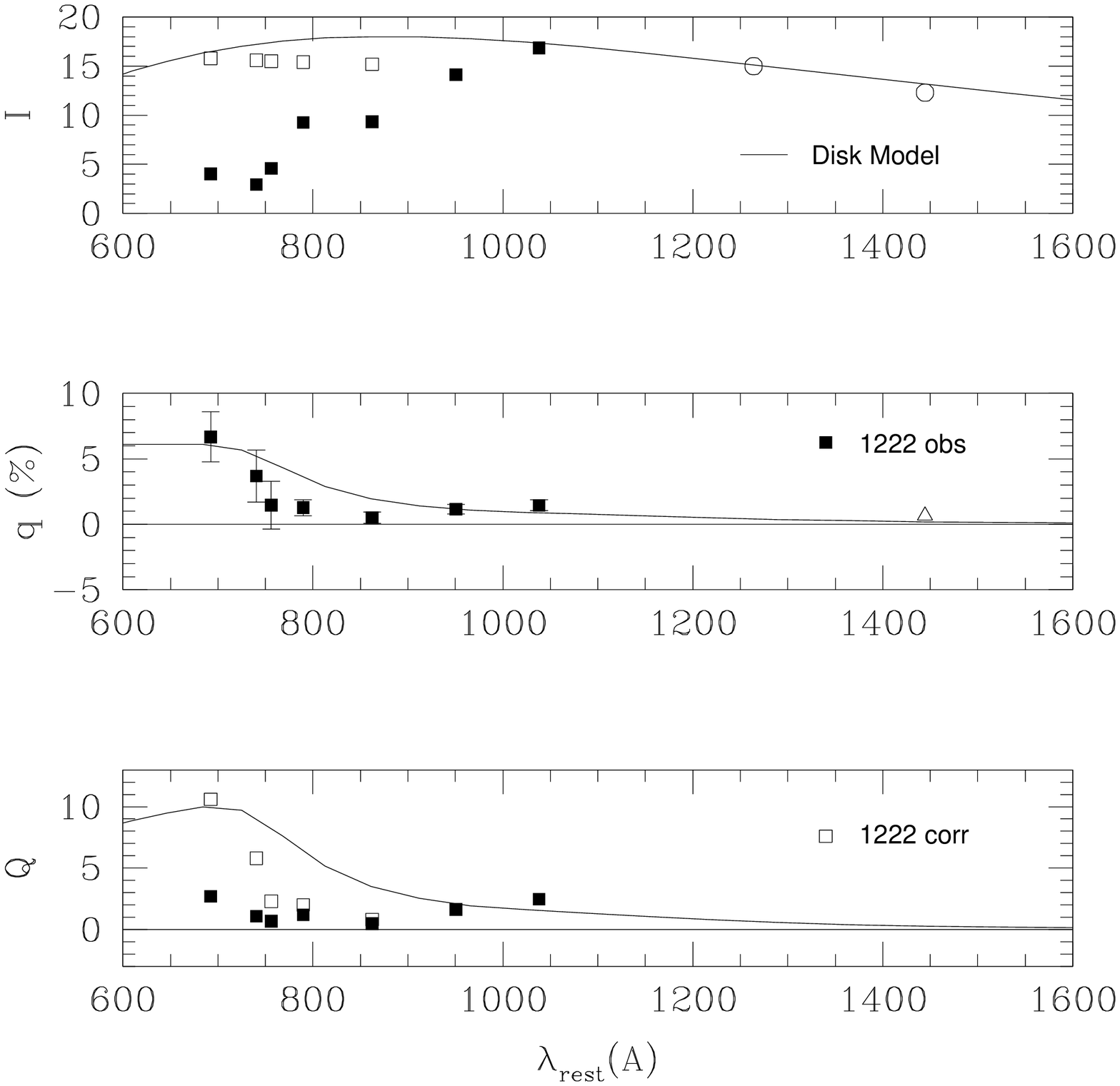]{Accretion disk model compared with 
observations of PG 1222+228 (see text).  
Observations are from Impey \etal\ 1995 (squares), Bechtold \etal\
1984 (open circles),  and
Webb \etal\ 1993 (open triangle).
Model disk with $a_* = 0.5$,
$M_9$ = 8.8 and \mdoto = 86 is viewed at angle \muobs = 0.25. 
Axes are same as in Fig. 2. \label{fig5}}

\clearpage

\evensidemargin -0.20in
\oddsidemargin -0.20in

\begin{deluxetable}{crrrrrrrr}

\tablenum{1}
\tablewidth{6.8in}
\tablecaption{Continuum Polarization of PG 1222+228$^{a}$}
\tablecolumns{9}
\tablehead{
\colhead{Grating}
& \colhead{$\lambda_{min}$}
& \colhead{$\lambda_{max}$}
& \colhead{ $I$}
& \colhead{$q(\%)$ }
& \colhead{$u(\%)$}
& \colhead{$p(\%)$}
& \colhead{$\theta$}
& \colhead{$q^\prime$}
}

\startdata

G190H  &1994 &2224  &4.04$\pm$0.04  &5.8$\pm$1.8  
&-3.4$\pm$2.6  &6.4$\pm$2.0  &165$\pm$16 &6.7$\pm$1.9\nl

Both &2224 &2287  &2.94$\pm$0.04  &2.4$\pm$2.0
&-3.8$\pm$2.1   &4.0$\pm$2.0  &151$\pm$26 &3.7$\pm$2.0\nl

Both &2287 &2319  &4.59$\pm$0.05   &0.2$\pm$1.8  
&-3.4$\pm$1.9   &2.9$\pm$1.9  &136$\pm$31 &1.5$\pm$1.8 \nl

G270H  &2319 &2492  &9.27$\pm$0.04  &1.0$\pm$0.6 
&-1.0$\pm$0.6   &1.2$\pm$0.6  &157$\pm$26 &1.3$\pm$0.6 \nl

G270H  &2492 &2761  &9.33$\pm$0.02  &0.5$\pm$0.4 
&-0.0$\pm$0.4   &0.3$\pm$0.4  &179$\pm$48  &0.5$\pm$0.4 \nl

G270H  &2761 &3029  &14.14$\pm$0.03   &0.9$\pm$0.4  
&-0.8$\pm$0.4  &1.2$\pm$0.4  &159$\pm$17  &1.2$\pm$0.4 \nl

G270H  &3029 &3295  &16.85$\pm$0.04   &1.7$\pm$0.4  
&0.2$\pm$0.4   &1.6$\pm$0.4  &4$\pm$14  &1.5$\pm$0.4 \nl

\enddata

\tablenotetext{{\it a}}{Data of Impey et al. (1995, 1996) rebinned with
respect to the apparent LLS at $\lambda$2300.
Wavelength limits (observed) are given in \AA.
Stokes parameters are given as $I$, the measured flux
(units 10$^{-16}$ erg
cm$^{-2}$ s$^{-1}$ \AA$^{-1}$), $q\equiv Q/I$, and $u\equiv U/I$.
The quantity 
$q^\prime$ is $Q/I$ rotated to P.A. 168$^{\circ}$.  Quoted uncertainties
are 1$\sigma$, derived from uncertainties in individual pixel
measurements.  Polarization $p$ is corrected for bias (Wardle \&
Kronberg 1974, eq. A3).
}

\end{deluxetable}

\clearpage

\evensidemargin -0.0in
\oddsidemargin  -0.0in
  
\plotone{fig1.eps}
\clearpage
\plotone{fig2.eps}
\clearpage
\plotone{fig3.eps}
\clearpage
\plotone{fig4.eps}
\clearpage
\plotone{fig5.eps}
\clearpage


\begin{thebibliography} {}

\bibitem {} Antonucci, R. R. J., Kinney, A. L., \& Ford, H. C. 1989, \apj, 342, 64

\bibitem {} Arav, N.
1997, in ``Mass Ejection from AGN'', ed. N. Arav, I. Shlosman, \& R. J. Weymann, A.S.P.
Conf. Ser., Vol. 128, p. 208

\bibitem {} Arav, N., Korista, K. T., de Kool, M., Junkkarinen, V. T., \& Begelman, M.
1999, \apj, 516, 27

\bibitem {} Arav, N., Shlosman, I., \& Weymann, R. J. (eds.) 1997, ``Mass Ejection
from AGN'', A.S.P. Conf. Ser., Vol. 128.

\bibitem {} Barlow, T. A., Hamann, F., \& Sargent, W. L. W. 
1997, in ``Mass Ejection from AGN'', ed. N. Arav, I. Shlosman, \& R. J. Weymann, A.S.P.
Conf. Ser., Vol. 128, p. 13

\bibitem {} Bechtold, J. \etal\ 1984, \apj, 281, 76

\bibitem {} Beloborodov, A. M., \& Poutanen, J. 1999, \apjl, 517, L77

\bibitem {} Blaes, O., \& Agol, E. 1996, \apjl, 369, L41

\bibitem {} Blaes, O., \& Shields, G. A. 1999, unpublished

\bibitem {} Brandt, W. N., Laor, A., \& Wills, B. J. 1999, \apj, in press
 (astro-ph/9908016)

\bibitem {} Chandrasekhar, S. 1960, Radiative Transfer (New York: Dover)

\bibitem {} Cohen, M. L., Ogle, P. M., Tran, H. D., Vermuellen, R. C., Miller, J. S.,
Goodrich, R. W., \& Martel, A. R. 1995, \apjl, 448, L77

\bibitem {} Ganguly, R., Churchill, C. W., \& Charlton, J. C. 1998, \apjl, 498, L103

\bibitem {} Goodrich, R. R., \& Miller, J. S. 1995, \apjl, 448, L73

\bibitem {} Hamann, F. 1997, \apjs, 109, 279

\bibitem {} Hamann, F., Barlow, T., Cohen, R. D., Junkkarinen, V., \& Burbidge, E. M.
1997, in ``Mass Ejection from AGN'', ed. N. Arav, I. Shlosman, \& R. J. Weymann, A.S.P.
Conf. Ser., Vol. 128, p. 19

\bibitem {} Hamann, F., \& Ferland, G. 1993, \apj, 418, 11

\bibitem {} Hines, D. C., \& Wills, B. J. 1995, \apjl, 448, L69

\bibitem {} Impey, C., Malkan, M., Webb, W., \& Petry, C. 1995,
 \apj, 440, 80

\bibitem {} Impey, C.,  \&  Petry, C. 1999, personal communication

\bibitem {} Impey, C.,  Petry, C., Malkan, M., \& Webb, W. 1996,
 \apj, 463, 473

\bibitem {} Junkkarinen, V., Beaver, E. A., Burbidge, E. M., Cohen,
R. D., Hamann, F., \&  Lyons, R. W. 1997, in ``Mass Ejection from AGN'', ed. N.
Arav, I. Shlosman, \& R. J. Weymann, A.S.P. Conf. Ser., Vol. 128, p. 19

\bibitem {} Koratkar, A., Antonucci, R., Goodrich, R.
 Bushouse, H., \& Kinney, A. 1995, \apj, 450, 501.

\bibitem {} Koratkar, A., Antonucci, R., Goodrich, R., \& Storrs, A. 1998, \apj, 503,
599

\bibitem {} Koratkar, A., \& Blaes, O. 1999, \pasp, 111, 1

\bibitem {} Koratkar, A., Kinney, A. L., \& Bohlin, R. C. 1992, 
\apj, 400, 435

\bibitem {} Krolik, J. H., \& Begelman, M. C. 1986, \apjl, 308, L55

\bibitem {} Laor, A., \& Draine, B. T. 1993, \apj, 402, 441

\bibitem {} Laor, A., Netzer, H., \& Piran, T. 1990, 
\mnras, 242, 560

\bibitem {} Lee, H.-W., \& Blandford, R. D. 1997, \mnras, 288, 19

\bibitem {} Mushotzky, R. F. 1999, personal communication

\bibitem {} Ogle, P. 1997, in ``Mass Ejection
from AGN'', ed. N. Arav, I. Shlosman, \& R. J. Weymann, A.S.P. Conf.
Ser., Vol. 128, p. 78

\bibitem {} Ogle, D. C.,  Cohen, M. H.,  Miller, J. S., Tran, H. D.,
 Goodrich, R. W., Martel, A. R 1999, \apjs, 125, 1

\bibitem {} Richards, G. T., York, D. G., Yanny, B., Kollgaard, R. I.,
Laurent-Muehleisen, S. A., \& Vanden Berk, D. E. 1999, \apj, 513, 576

\bibitem {} Sargent, W. L. W., Steidel, C. C., \& Boksenberg, A. 1988, 
  \apjs, 68, 539

\bibitem {} Sargent, W. L. W., Steidel, C. C., \& Boksenberg, A. 1989,
  \apjs, 69, 703

\bibitem {} Schmidt, M., \& Green, R. F. 1983, \apj, 269, 352

\bibitem {} Schmidt, G. D., \&  Hines, D. C. 1999, \apj, 512, 125

\bibitem {} Shields, G. A. 1997, in ``Mass Ejection
from AGN'', ed. N. Arav, I. Shlosman, \& R. J. Weymann, A.S.P. Conf.
Ser., Vol. 128, p. 214

\bibitem {} Shields, G. A., Wobus, L., \& Husfeld, D. 1998, \apj, 496, 743

\bibitem {} Steidel, C. C., \& Sargent, W. L. W., 1992, \apjs, 80, 1

\bibitem {} Stockman, H. S., Moore, R. L., \& Angel, J. R. P. 1984,
\apj, 279, 485

\bibitem {} Telfer, R. C., Kriss, G. A., Zheng, W., Davidsen, A. F., \& 
  Green, R. 1999, \apj, 509, 132

\bibitem {} Turnshek, D. A., Kopko, M., Jr., Monier, E., Noll, D., 
Espey, B. R., \& Weymann, R. J. 1996, \apj, 463, 110

\bibitem {} Wampler, E. J., \& Ponz, D. 1985, \apj, 298, 448.

\bibitem {} Wardle, J. F. C. \& Kronberg, P. P. 1974, \apj, 194, 254

\bibitem {} Webb, W., Malkan, M., Schmidt, G., \& Impey, C., 1993, 
\apj, 419, 494

\bibitem {} Weymann, R. J. 1997, in ``Mass Ejection
from AGN'', ed. N. Arav, I. Shlosman, \& R. J. Weymann, A.S.P. Conf.
Ser., Vol. 128, p. 3

\bibitem {} Weymann, R. J., Morris, S. L., Foltz, C. B., 
\& Hewett, P. C. 1991, \apj, 373, 23

\bibitem {} Wilkes, B. J., Tananbaum, H., Worrall, D. M., Avni, Y., Oey, M. S., \&
Flanagan, J. 1994, \apjs, 92, 53

\bibitem {} Zheng, W., Kriss, G., Telfer, R. C., Grimes, J. P., Davidsen, \& A. F. 
1997, \apj, 475, 469

\end{thebibliography}
\end{document}